\documentclass[a4paper,oneside]{article}

\usepackage{microtype}
\usepackage[sc]{mathpazo}
\usepackage[T1]{fontenc}
\usepackage[DIV=11,BCOR=0cm,headinclude=true]{typearea}

\usepackage{amsfonts}
\usepackage{amsmath}
\usepackage{amssymb}
\usepackage{amsthm}
\usepackage{graphicx}
\usepackage{float}
\usepackage[config,textfont=sl]{caption,subfig}
\usepackage[british]{babel}
\usepackage{bm}

\usepackage[maxbibnames=100,autocite=superscript,style=custom-nature,sorting=none,defernumbers=true]{biblatex}

\usepackage{authblk}

\newcommand{\bv}[1]{\boldsymbol{#1}}
\newcommand{\bra}[1]{\langle{#1}|}
\newcommand{\ket}[1]{|{#1}\rangle}
\newcommand{\braket}[2]{\langle{#1}|{#2}\rangle}
\newcommand{\tensor}[1]{\bar{\bar{#1}}}

\DeclareMathOperator*{\real}{Re}
\DeclareMathOperator*{\imag}{Im}

\author[1,2,$\dagger$]{Parry Y.\ Chen}
\author[1]{David J.\ Bergman}
\author[2]{Yonatan Sivan}
\affil[1]{School of Physics and Astronomy, Raymond and Beverly Sackler Faculty of Exact Sciences, Tel Aviv University, Israel}
\affil[2]{Unit of Electro-optic Engineering, Ben-Gurion University, Israel}
\affil[$\dagger$]{\it{parryyu@post.bgu.ac.il}}

\title{Generalizing normal mode expansion of electromagnetic Green's tensor to lossy resonators in open systems}
\date{\today}

\begin{document}
\maketitle

\begin{abstract} \noindent 
We generalize normal mode expansion of Green's tensor $\tensor{G}(\bv{r},\bv{r}')$ to lossy resonators in open systems, resolving a longstanding open challenge. We obtain a simple yet robust formulation, whereby radiation of energy to infinity is captured by a complete, discrete set of modes, rather than a continuum. This enables rapid simulations by providing the spatial variation of $\tensor{G}(\bv{r},\bv{r}')$ over both $\bv{r}$ and $\bv{r}'$ in one simulation. Few eigenmodes are often necessary for nanostructures, facilitating both analytic calculations and unified insight into computationally intensive phenomena such as Purcell enhancement, radiative heat transfer, van der Waals forces, and F\"{o}rster resonance energy transfer. We bypass all implementation and completeness issues associated with the alternative quasinormal eigenmode methods, by defining modes with permittivity rather than frequency as the eigenvalue. We obtain true stationary modes that decay rather than diverge at infinity, and are trivially normalized. Completeness is achieved both for sources located within the inclusion and the background through use of the Lippmann-Schwinger equation. Modes are defined by a linear eigenvalue problem, readily implemented using any numerical method. We demonstrate its simple implementation on COMSOL Multiphysics, using the default inbuilt tools. Results were validated against direct scattering simulations, including analytical Mie theory, attaining arbitrarily accurate agreement regardless of source location or detuning from resonance.
\end{abstract}

\section{Introduction}
Green's functions $G(\bv{r},\bv{r}')$ and Green's tensors $\tensor{G}(\bv{r},\bv{r}')$ are essential tools for solving linear wave equations with source terms, being the fundamental solution of a unit impulse. Closed form expressions of $\tensor{G}(\bv{r},\bv{r}')$ are known for free space and a limited number of simple geometries, but in a general geometry such as an arbitrarily-shaped cavity, it has a non-trivial variation over $\bv{r}$ and $\bv{r}'$.\autocite{morse1946methods,arfken1999mathematical,novotny2012principles} Numerical simulation is often the only available option, but the computational burden becomes prohibitive if repeated simulations are required for every source position and orientation. 

A more palatable alternative is to expand $\tensor{G}(\bv{r},\bv{r}')$ in the basis of the cavity's eigenmodes, which can be obtained from a single simulation. These provide the full spatial variation of $\tensor{G}(\bv{r},\bv{r}')$ over $\bv{r}$ and $\bv{r}'$, especially since only a few eigenmodes are often necessary for expansion. Modal expansion of $\tensor{G}(\bv{r},\bv{r}')$ has a long history for conservative systems, such as closed cavities without loss.\autocite{morse1946methods} Here, the eigenmodes are \emph{normal modes}, which are stationary states with real eigenfrequencies. These provide a complete and orthogonal basis, with completeness being necessary for a valid expansion of $\tensor{G}(\bv{r},\bv{r}')$. Less straightforward is modal expansion for lossy resonators in open systems, a topical problem on which research has intensified recently. Conceptually, the simplest generalization of modal expansion to handle losses is to permit the modes to have complex eigenfrequencies, yielding \emph{quasinormal modes}. \autocite{siegert1939derivation,baum1971singularity,baum1976emerging,lai1990time,kristensen2012generalized,doost2013resonant,sauvan2013theory} However, use of quasinormal modes carries a number of unwelcome side effects. 

We adopt a different approach in this paper, and instead seek a direct generalization of normal modes to lossy resonators in open systems, thus expanding $\tensor{G}(\bv{r},\bv{r}')$ in true stationary states, an open problem since the inception of normal mode expansions. We demonstrate that a generalized normal mode expansion (GENOME) recovers both the simplicity and rigor of normal modes observed in conservative systems. We specifically treat the electromagnetic Green's tensor, but also remark that GENOME can be applied to any other wave equation, such as for acoustics, elasticity,\autocite{kantor1984improved,cohen2003effective} quantum mechanical scattering,\autocite{band2013quantum} and linearized gravitational waves.\autocite{chandrasekhar1975quasi,leung1997quasinormal}

The electromagnetic Green's tensor $\tensor{G}(\bv{r},\bv{r}')$ is fundamental to the photonic density of states, relating both to the power radiated by a classical dipole antenna, and also the spontaneous emission rate of quantum emitters under a semi-classical treatment.\autocite{novotny2012principles,carminati2015electromagnetic} Green's tensor can be strongly influenced by appropriately designed nanostructures. A broad range of electromagnetic processes and quantum optics phenomena may be enhanced or altered, including emission from individual atoms and molecules, known as the Purcell effect, charge transfer between molecules, known as F\"{o}rster resonance energy transfer, as well as emission from bulk sources, such as in nonlinear wave mixing, radiative heat transfer, van der Waals forces, and quantum friction, among others. All such phenomena are linked to $\tensor{G}(\bv{r},\bv{r}')$, owing to its fundamental definition,
\begin{equation}
\nabla\times(\nabla\times \tensor{G}) - k^2\epsilon(\bv{r})\tensor{G} = \tensor{I}\delta^3(\bv{r}-\bv{r}'),
\label{eq:greenfn}
\end{equation}
which gives the electrodynamic response of a detector at $\bv{r}$, due to a point source at $\bv{r}'$ radiating at frequency $\omega = c k$, in the presence of an inclusion defined by $\epsilon(\bv{r})$, where $\tensor{I}$ is the unit tensor. 

$\tensor{G}(\bv{r},\bv{r}')$ can be obtained via numerical simulation by placing a radiating dipole at $\bv{r}'$. Finding the full spatial variation becomes a laborious task, since the location and orientation of the source dipole must be continuously varied, requiring repeated simulation. Instead, modal expansion provides\begin{equation}
\tensor{G}(\bv{r}, \bv{r}') = c^2\sum_m \frac{\bv{E}_m(\bv{r}) \otimes \bv{E}^*_m(\bv{r}')}{\omega_m^2 - \omega^2},
\label{eq:simple}
\end{equation}
for conservative systems.\autocite{morse1946methods,novotny2012principles,carminati2015electromagnetic,arfken1999mathematical} Here, $\omega_m$ are the eigenfrequencies of each mode, and $\otimes$ defines the outer product between two vectors. Thus, the fields produced by any point source is immediately available by simply evaluating the complex conjugate modal field $\bv{E}_m^*(\bv{r}')$ at $\bv{r}'$. The same eigenmode $\bv{E}_m(\bv{r})$ also gives variation over detector positions $\bv{r}$, thus providing unified analytical insight across a wide variety of optical phenomena. Indeed, often a single mode suffices, since the detuning $\omega_m^2-\omega^2$ from the other resonances is typically large when the nanostructure is small. All nine tensor components are available, which would otherwise require nine separate simulations for each $\bv{r}'$.\autocite{vos2009orientation} Finally, the total fields $\bv{E}(\bv{r})$ produced by bulk sources, or any arbitrary source distribution $\bv{J}(\bv{r}')$, can be calculated with the same ease: by superposing contributions from different source positions,
\begin{equation}
\bv{E}(\bv{r}) = i\omega\mu_0 \int \tensor{G}(\bv{r},\bv{r}') \cdot \bv{J}(\bv{r}')\, dV',
\label{eq:simpleJ}
\end{equation}
where $\tensor{G}(\bv{r}, \bv{r}')$ is expanded using \eqref{eq:simple}, yielding overlap integrals between $\bv{J}(\bv{r}')$ and $\bv{E}_m^*(\bv{r}')$.

The expansion \eqref{eq:simple} is adequate when losses are negligible. However, resonators with non-negligible losses have attracted significant research attention recently, particularly plasmonic nanoresonators, consisting for example of a metallic inclusion in a uniform background. Metal-dielectric interfaces produce sharp field concentrations, generating electromagnetic hotspots ideal for influencing density of states. However, energy is lost due to material absorption, and is continuously radiated into the background. These losses can be treated in a modal expansion by coupling to the material degrees of freedom,\autocite{bhat2006hamiltonian} and the continuum of background electromagnetic modes,\autocite{ching1998quasinormal} respectively. But the utility of \eqref{eq:simple} is diminished since the expansion ceases to be via a limited set of discrete electromagnetic modes.

Expansion via a discrete set is once again possible using quasinormal modes, yielding an expression similar to \eqref{eq:simple}.\autocite{lee1999dyadic,doost2013resonant,kristensen2014modes,sauvan2014modal,muljarov2016exact} Discreteness stems from imposing Sommerfeld radiation boundary conditions to account for radiation.\autocite{dai2012generalized} The imaginary part of the complex eigenfrequency relates to the finite lifetime or decay rate of energy to all loss channels. However, complex eigenfrequencies introduce their own implementational and interpretational difficulties, such as a difficult-to-solve non-linear eigenvalue problem, unphysical exponentially diverging fields, and the need to define permittivities at complex frequencies. More fundamentally, quasinormal modes only provide a complete expansion for source positions $\bv{r}'$ inside the resonator,\autocite{leung1994completeness,muljarov2016exact} giving inaccuracies for exterior sources. Nevertheless, quasinormal modes yield very satisfactory results for many practical applications, as much progress has been achieved in the past several years in their development, and in overcoming their previous limitations. See Section \ref{sec:quasi} for further details.

GENOME bypasses these complexities, providing many advantages, both fundamental and practical. Instead of using complex frequency modes to account for loss, true stationary modes are obtained by designating the permittivity of the nanostructure $\epsilon_m$ to be the complex eigenvalue. Thus, we fix a real $k$, and also fix the inclusion's geometry and background permittivity of the simulation domain, consequently determining the complex $\epsilon_m$ which brings the system to resonance. Our GENOME approach is based on the \emph{spectral decomposition} formalism first developed in a series of papers by Bergman and others for electrostatics,\autocite{bergman1979dielectric,bergman1979dielectrica} and then electrodynamics,\autocite{bergman1980theory,farhi2016electromagnetic} with similar methods having also been developed independently by others.\autocite{agranovich1999generalized,sandu2012eigenmode} In the electrostatic limit, the formalism has been used for applications such as spasers,\autocite{bergman2003surface} self-similar antennas,\autocite{li2003self} disordered media,\autocite{stockman2001localization,stockman2002coherent} and second harmonic generation,\autocite{stockman2004enhanced,li2005enhanced} as well as the computation of effective medium parameters, bounds on them and associated sum rules.\autocite{bergman1979dielectrica,bergman1992bulk,li2003self} In the electrodynamic limit, it has been used for spherical and Veselago-Pendry lens geometries,\autocite{bergman1980theory,farhi2016electromagnetic} and for the analysis of lifetimes and second harmonic generation,\autocite{bergman2005radiative,reddy2017revisiting} in particular, resolving long standing ambiguities on the definitions of boundary conditions for surface second harmonic generation.

Completeness is an important consideration, being necessary for successful modal expansion.\autocite{morse1946methods} Direct expansion using eigenmodes, including our eigenpermittivity modes, is insufficient to give completeness everywhere. Key to the success of GENOME is use of the Lippmann-Schwinger equation, which extends completeness from interior sources to all exterior sources $\bv{r}'$ and all detector locations $\bv{r}$.\autocite{bergman1980theory} Use of the Lippmann-Schwinger equation distinguishes GENOME from other methods that also employ eigenpermittivity modes.\autocite{ge2010steady} Thus, GENOME converges to the correct solution with arbitrary precision as more eigenmodes are considered, regardless of detuning from the inclusion's resonances and source or detector coordinates. Unlike \eqref{eq:simple}, our expansion also efficiently reproduces the divergence and non-transverse components of $\tensor{G}(\bv{r},\bv{r}')$ at the source, without requiring many modes. 

Modal completeness affords great practical utility. Crucially, modes of analytically insoluble systems can be generated effortlessly and reliably either using modes of a simpler system or even an analytically soluble system as a basis. For example, since quasinormal modes are complete internally, modes of a sphere can generate modes of wedges or any arbitrary shape enclosed by the sphere,\autocite{doost2014resonant} such as split ring resonators. A compelling benefit of GENOME over quasinormal modes is that the modes of clusters and arrays of nanostructures can be obtained from known modes of the constituents without further numerical simulation, for example obtaining dimer modes from monomer modes.\autocite{bergman1980theory,bergman1979dielectrica} This constitutes a generalization of the celebrated theory of linear combination of molecular orbitals to electromagnetic structures,\autocite{mulliken1967spectroscopy,huheey1997inorganic} and provides a rigorous generalization of an approximate hybridization approach developed in the context of nanoplasmonics.\autocite{prodan2003hybridization} 

Our generalized normal modes remain well-behaved throughout all space, since frequency and background permittivity remain real and physical. In particular, our eigenmodes both decay to zero at infinity and satisfy the vector equivalent of the Sommerfeld radiation condition, the Silver-M\"{u}ller condition. In contrast to quasinormal modes, this corresponds to the physical far-field solution and yields the correct energy radiated into the background. Consequently, normalization of modes is also rapid and trivial, achieved by a volume integral over the inclusion's interior.

Our modes are always generated by a linear eigenvalue equation, even when material dispersion is present. Thus, eigenmode search is simple and readily automated, with no need for delicate root searches in the complex plane, instead relying on the many powerful, robust linear algebra algorithms. We demonstrate its straightforward general implementation on COMSOL, a commercially available finite element package, by adapting its inbuilt eigenfrequency solver with a simple substitution trick. Due to the simplicity of the linear eigenvalue equation, implementation via any method capable of producing eigenmodes is possible, including volume integral methods, discrete dipole approximation, and plane wave expansion. Analytic solutions are also available for simple spherical, cylindrical, and planar geometries.\autocite{bergman1980theory,farhi2016electromagnetic,chen2017robust} 

The paper is organized as follows. In Section \ref{sec:method}, we develop GENOME for lossy, open systems. The exposition is relatively self-contained, but technical proofs are omitted, focusing instead on conveying underlying insight into the properties of our method. Special attention is given to the generalized normal modes in Section \ref{sec:eigenmodes}. In Section \ref{sec:quasi}, we offer a comprehensive comparison between GENOME and the alternative quasinormal mode expansion. Section \ref{sec:numerics} details the brief numerical implementation on COMSOL Multiphysics, with numerical examples that demonstrate the completeness of GENOME. Further discussion on the properties of eigenpermittivity modes is also provided. Section \ref{sec:conc} presents a summary and conclusion.

\section{Generalized Normal Mode Expansion}
\label{sec:method}
\subsection{Lippmann-Schwinger equation}
\label{sec:lippsch}
The foundation of GENOME is the Lippmann-Schwinger equation for electrodynamics, which is also the basis of two families of related numerical schemes: method of moments (MoM) and discrete dipole approximation (DDA), also known as volume integral or coupled dipole methods, respectively.\autocite{purcell1973scattering,harrington1993field,lakhtakia1990macroscopic,novotny2012principles} In quantum mechanics, the Lippmann-Schwinger equation is used for scattering calculations, and is commonly solved using the Born approximation or Born Series. 

The Lippmann-Schwinger equation is obtained from Maxwell's equations,
\begin{equation}
\nabla\times(\nabla\times\bv{E}) - k^2\epsilon(\bv{r})\bv{E} = i\omega\mu_0\bv{J},
\label{eq:maxwell}
\end{equation}
assuming harmonic $e^{-i\omega t}$ time variation and non-magnetic media. We begin by assuming that the structure defined by its permittivity profile $\epsilon(\bv{r})$ rests in a background of uniform permittivity $\epsilon_b$. This permits the manipulation of $\eqref{eq:maxwell}$ to yield
\begin{equation}
\nabla\times(\nabla\times\bv{E}) - k^2\epsilon_b\bv{E} = i\omega\mu_0\bv{J} + k^2(\epsilon(\bv{r})-\epsilon_b)\bv{E}.
\label{eq:inhomowave}
\end{equation}
In \eqref{eq:maxwell}, notice that $\bv{J}$, the imposed free current source, sits alone on the right hand side (RHS) while the response of the inclusion is on the LHS. In \eqref{eq:inhomowave}, both $\bv{J}$ and the response of the inclusion are on the RHS, and the second term can be interpreted as the bound currents produced by the inclusion.

Since the operator on the LHS of \eqref{eq:inhomowave} is no longer a function of $\bv{r}$, \eqref{eq:inhomowave} can be solved using the simple Green's function for uniform media,
\begin{equation}
\nabla\times(\nabla\times\tensor{G}_0) - k^2\epsilon_b\tensor{G}_0 = \tensor{I}\delta^3(\bv{r}-\bv{r}'),
\label{eq:greensfree}
\end{equation}
which has a simple known analytic form $\tensor{G}_0(|\bv{r}-\bv{r}'|)$ depending on the dimensionality of the problem.\autocite{arfken1999mathematical} Applying \eqref{eq:greensfree} to both terms on the RHS of \eqref{eq:inhomowave} yields its Green's function solution, which is the desired Lippmann-Schwinger equation, 
\begin{equation}
\bv{E}(\bv{r}) = \bv{E}_0(\bv{r}) + k^2 \int \tensor{G}_0 (|\bv{r} - \bv{r}'|) (\epsilon(\bv{r}') - \epsilon_b) \bv{E}(\bv{r}')\, d\bv{r}'.
\label{eq:lippsch}
\end{equation}
The term $\bv{E}_0(\bv{r})$ is the known radiation pattern of external sources in a uniform background
\begin{equation}
\bv{E}_0(\bv{r}) = i\omega\mu_0 \int \tensor{G}_0(|\bv{r} - \bv{r}'|) \bv{J}(\bv{r}')\, d\bv{r}',
\label{eq:E0}
\end{equation}
which is simple to solve since it is independent of the inclusion. In particular, standard textbook expressions are available for the $\bv{E}_0(\bv{r})$ of point dipoles and other simple source configurations.\autocite{jackson2007classical}

Crucial to the Lippmann-Schwinger equation is that the same simple Green's tensor appears in both \eqref{eq:lippsch} and \eqref{eq:E0}, a property exploited by all resultant methods including GENOME. As mentioned, the Lippmann-Schwinger equation, \eqref{eq:lippsch}, can be solved numerically. This involves spatial discretization, recasting \eqref{eq:lippsch} in linear algebra form, which sometimes requires iterative solution until convergence.\autocite{martin1994iterative} This is due to the implicit nature of \eqref{eq:lippsch}, with the desired solution $\bv{E}(\bv{r})$ appearing inside the integral, thus forming a Fredholm integral equation of the second kind. The Lippmann-Schwinger equation can also be expanded in terms of basis functions, such as the cylindrical harmonic functions, again yielding a linear algebra problem, but obviating the need for iterative solution.\autocite{kristensen2010light}

We use the Lippmann-Schwinger equation as the foundation of a yet more powerful analytic method. Instead of solving \eqref{eq:lippsch} directly, we first find its stationary or self-sustaining solutions, corresponding exactly to the eigenpermittivity modes of the inclusion. By projecting onto this basis, we obtain an analytic solution of the Lippmann-Schwinger equation \eqref{eq:lippsch}. We then obtain the desired eigenmode expansion of \eqref{eq:greenfn} which, unlike \eqref{eq:greensfree}, may be regarded as the ``dressed Green's tensor''. The solution is rigorous and valid everywhere even though the eigenmodes form a complete set only in the interior. This follows from a key property of the Lippmann-Schwinger equations, enabling $\bv{E}(\bv{r})$ to be determined everywhere from knowledge of $\bv{E}(\bv{r})$ inside the inclusion only, where $(\epsilon(\bv{r})-\epsilon_b)$ is non-zero.

\subsection{The generalized normal modes}
\label{sec:eigenmodes}
To proceed with GENOME, we define the appropriate normal modes of the system. We also outline some key properties, from which many of the advantages of the method stem. The eigenvalue equation is obtained by neglecting $\bv{E}_0$ in \eqref{eq:lippsch}. At this point, we simplify the formulation by assuming that the permittivity of the inclusion is uniform, so that the simulation domain is defined by only two permittivities: an eigenpermittivity $\epsilon_m$ applicable to the inclusion's interior, and the fixed background $\epsilon_b$. This yields the simplified eigenvalue equation
\begin{equation}
s_m \bv{E}_m(\bv{r}) = k^2 \int \tensor{G}_0 (|\bv{r} - \bv{r}'|; k^2) \theta(\bv{r}') \bv{E}_m(\bv{r}')\, d\bv{r}',
\label{eq:eigen}
\end{equation}
where $s_m$ is the $m$th eigenvalue, known as the Bergman spectral paramater,\autocite{stockman2011spaser}
\begin{equation}
\frac{1}{s_m} \equiv \epsilon_m - \epsilon_b,
\label{eq:eigenvalue}
\end{equation}
$\theta(\bv{r})$ is a step function that is unity inside the inclusion and zero elsewhere, and the dependence of the Green's tensor on $k^2$ is explicitly noted. 

The eigenvalue in \eqref{eq:eigen} and \eqref{eq:eigenvalue} is $\epsilon_m$, representing the inclusion permittivity, which contrasts with the more prevalent choice, $k$. In other words, $k$ is held fixed while $\epsilon_m$ is varied until the inclusion is at resonance. As a simple example, the modes of a sphere can be determined by the poles of its scattering matrix, which for quasistatic fields varies as $(\epsilon_i-\epsilon_b)/(\epsilon_i+2\epsilon_b)$.\autocite{bohren1983absorption} Thus, the eigenpermittivity at long wavelengths is $\epsilon_m = -2\epsilon_b$. By specifying $k$ to be real, stationary modes are obtained. Moreover, the free-space Green's tensor decays appropriately rather than diverges as $\bv{r} \rightarrow \infty$, as do the resulting eigenmodes. Note that this choice also leads to a linear eigenvalue problem, since $\tensor{G}_0 (|\bv{r} - \bv{r}'|; k^2)$ is not a function of $\epsilon_m$. 

The modes in general have complex eigenvalues $\epsilon_m$, with the imaginary part corresponding to gain across the interior of the inclusion. This gain compensates for the energy lost from the inclusion due to radiation, permitting the mode to remain in a stationary state. Thus, the complex eigenpermittivity carries a simple interpretation, being the lasing threshold of the inclusion at a particular frequency.\autocite{ge2010steady,stockman2011spaser} Despite the similarity between \eqref{eq:lippsch} and \eqref{eq:eigen}, the eigenvalues $\epsilon_m$ are in general unrelated to the actual permittivity of the inclusion to be solved in \eqref{eq:lippsch}. The actual permittivity is specified later in \eqref{eq:E0sol}, with the modes $\bv{E}_m$ serving as a complete orthonormal mathematical basis.\autocite{bergman1980theory} The only information from \eqref{eq:lippsch} that remains in \eqref{eq:eigen} is the frequency, and the geometry of the inclusion captured by $\theta(\bv{r})$. In that sense, our formulation separates the material properties from geometric properties, so the imaginary part of $\epsilon_m$ only needs to account for radiation losses experienced by the geometry. Indeed, the magnitude of $\imag(\epsilon_m)$ can be used to quickly identify whether a mode is bright or dark. Furthermore, the eigenmodes are applicable to any uniform inclusion permittivity, even complex permittivities, without modification. 

In passing, we mention that the restriction to inclusions with a uniform isotropic permittivity in \eqref{eq:eigen} is not necessary. Generalizations to non-uniform and anisotropic inclusions are possible. The eigenmodes would still be defined by \eqref{eq:eigen}, but $\theta(\bv{r})$ would no longer be a step function, requiring generalization to reflect the $\tensor{\epsilon}(\bv{r})$ profile to be simulated. The eigenvalues are still $s_m$, but their interpretation as the eigenpermittivities of the inclusion \eqref{eq:eigenvalue} would no longer be valid. Lastly, non-uniform or anisotropic background media are possible, by using the relevant Green's tensor in place of $G_0(|\bv{r}-\bv{r}'|)$.\autocite{chen1983theory,cortes2017super}

\subsection{Expansion via normal modes}
\label{sec:expand}
We take as given that the relatively simple task of finding the radiation pattern in the uniform background $\bv{E}_0$ in \eqref{eq:E0} is complete. The final stage of GENOME is to solve the Lippmann-Schwinger equation \eqref{eq:lippsch} by using its normal modes, \eqref{eq:eigen}, to expand the source $\bv{J}(\bv{r})$. We largely follow the derivation of Ref.\ \parencite{bergman1980theory}. For notational brevity, we begin by casting the Lippmann-Schwinger equation \eqref{eq:lippsch} in operator form,
\begin{equation}
\bv{E} = \bv{E}_0 + u\hat{\Gamma}\hat{\theta}\bv{E},
\label{eq:lippschgamma}
\end{equation}
where $u$ now describes the permittivity of the actual inclusion $\epsilon_i$, 
\begin{equation}
u \equiv \epsilon_i-\epsilon_b.
\end{equation}
$\hat{\Gamma}$ is an integral operator incorporating the Green's function along with $k$, and $\hat{\theta}$ is the operator form of $\theta(\bv{r})$ which zeros the field outside the inclusion, so
\begin{equation}
\hat{\Gamma}\hat{\theta}\bv{E} \equiv k^2 \int \tensor{G}_0(|\bv{r} - \bv{r}'|) \theta(\bv{r}') \bv{E}(\bv{r}')\, d\bv{r}'.
\label{eq:gammadef}
\end{equation}

The formal solution to \eqref{eq:lippschgamma} is
\begin{equation}
\bv{E} = \frac{1}{1-u\hat{\Gamma}\hat{\theta}}\bv{E}_0.
\label{eq:formal}
\end{equation}
Note that in spectral theory, the operator $(1-u\hat{\Gamma}\hat{\theta})^{-1}$ in \eqref{eq:formal} is known as the resolvent.\autocite{cohen-tannoudji1998atom} Our solution for the unknown field $\bv{E}$ proceeds by projecting the known $\bv{E}_0$ on to the known normal modes $\bv{E}_m$. Specifically, we define the projection operator $\hat{I}$, which in bra-ket notation is
\begin{equation}
\hat{I} = \sum_m \hat{\theta}\ket{E_m}\bra{E_m}\hat{\theta}.
\label{eq:project}
\end{equation}
This simple form is valid because the modes obey a biorthogonality relation.\autocite{bergman1980theory} By including $\hat{\theta}$ in $\hat{I}$, we expand only over the interior fields. This avoids an unwieldy integral over all space, and also expands only in the region where the eigenmodes provide a complete basis. Note that this projection operator assumes that the modes are normalized, 
\begin{equation}
\bra{E_m}\hat{\theta}\ket{E_m} = 1.
\label{eq:norm}
\end{equation}
The unknown field $\ket{E}$ is then
\begin{equation}
\hat{\theta}\ket{E} = \sum_m \hat{\theta}\ket{E_m}\bra{E_m}\frac{\hat{\theta}}{1-u\hat{\Gamma}\hat{\theta}}\ket{E_0}.
\label{eq:inteqn}
\end{equation}

Next is the key step of GENOME. Instead of applying the operator $(1-u\hat{\Gamma}\hat{\theta})^{-1}$ to $\ket{E_0}$, which would result in a lengthy numerical calculation via the Born series, we exploit the freedom offered by the unified nature of the Green's function in \eqref{eq:E0} and \eqref{eq:eigen} to operate on $\bra{E_m}$ instead, immediately yielding an exact analytic solution. We invoke the adjoint form of eigenvalue equation \eqref{eq:eigen},
\begin{equation}
\bra{E_m}\hat{\theta}\hat{\Gamma} = \bra{E_m}s_m.
\label{eq:eigenadj}
\end{equation}
It is critical here that the eigenmodes \eqref{eq:eigen} share a predefined frequency equivalent to the desired $k$ of the Lippmann-Schwinger equation \eqref{eq:lippsch} to be solved. This enables the freedom to interchange, since the Green's tensor represented by $\hat{\Gamma}$ in \eqref{eq:inteqn} and \eqref{eq:eigenadj} are identical. This obtains from \eqref{eq:inteqn} the total interior field $\hat{\theta}\ket{E}$,
\begin{equation}
\hat{\theta}\ket{E} = \sum_m \hat{\theta}\ket{E_m} \frac{1}{1-u s_m} \bra{E_m}\hat{\theta}\ket{E_0},
\label{eq:intsol}
\end{equation}
expressed in terms of overlap integrals. 

To obtain the fields everywhere, \eqref{eq:intsol} is inserted into the original Lippmann-Schwinger equation \eqref{eq:lippschgamma}, this time operating $\hat{\Gamma}\hat{\theta}$ on $\ket{E_m}$ to give
\begin{equation}
\ket{E} = \ket{E_0} + \sum_m \ket{E_m} \frac{u s_m}{1-u s_m} \bra{E_m}\hat{\theta}\ket{E_0}.
\label{eq:E0solus}
\end{equation}
Thus, with the aid of the Lippmann-Schwinger equation, we have obtained an expansion valid over all space even though we only expanded the fields inside the inclusion. For convenience, \eqref{eq:E0solus} can be expressed explicitly in terms of permittivities,
\begin{equation}
\ket{E} = \ket{E_0} + \sum_m \ket{E_m} \frac{\epsilon_i - \epsilon_b}{\epsilon_m-\epsilon_i} \bra{E_m}\hat{\theta}\ket{E_0}.
\label{eq:E0sol}
\end{equation}

Equation \eqref{eq:E0sol} expresses the total fields of the system in terms of the radiation of the source in a uniform medium, with additional contributions from modes of the inclusion that are excited. The weight of each eigenmode is determined in part by the detuning between the inclusion permittivity, $\epsilon_i$, and the eigenmode, $\epsilon_m$. The eigenmode with the most similar permittivity is the dominant contributor to the radiated energy, and the series converges rapidly onto the true solution. Secondly, the electrodynamic interaction between the source and the inclusion is entirely encoded within the geometric factor $\bra{E_m}\hat{\theta}\ket{E_0}$, representing the spatial overlap between the incident field and the mode being excited. The explicit form of this overlap integral is presented in Appendix \ref{sec:adjoint}. The solution \eqref{eq:E0sol} is exact up to truncation in $m$, since the Born series was avoided in obtaining \eqref{eq:intsol}, and arbitrary accuracy is possible by increasing $m$. The one set of eigenmodes $\ket{E_m}$ is applicable to all possible inclusion permittivities and excitations $\ket{E_0}$, the latter requiring only the evaluation of the overlap integral, which represents a small fraction of the total simulation time. 

The solution \eqref{eq:E0sol} is the most suitable form when the source is in the far field, so $\ket{E_0}$ has a known form, such as a plane wave or a beam. If however the source is in the near field, a second formulation is more convenient, expressed directly in terms of sources $\bv{J}(\bv{r})$.\autocite{farhi2016electromagnetic} This begins by casting \eqref{eq:E0} into operator form, yielding
\begin{equation}
\ket{E_0} = \frac{i}{\omega\epsilon_0} \hat{\Gamma} \ket{J}.
\end{equation}
After inserting into \eqref{eq:E0solus}, we obtain
\begin{equation}
\ket{E} = \ket{E_0} + \frac{i}{\omega\epsilon_0} \sum_m \ket{E_m} \frac{u s_m}{1-u s_m} \bra{E_m}\hat{\theta}\hat{\Gamma}\ket{J}.
\end{equation}
Again, by applying the operator $\hat{\theta}\hat{\Gamma}$ to $\bra{E_m}$ via \eqref{eq:eigenadj} rather than $\ket{J}$, a simple solution is obtained 
\begin{equation}
\ket{E} = \ket{E_0} + \frac{i}{\omega\epsilon_0} \sum_m \ket{E_m} \frac{u s_m^2}{1-u s_m} \braket{E_m}{J}.
\label{eq:Jevform}
\end{equation}
In terms of permittivities, \eqref{eq:Jevform} can be rewritten as
\begin{equation}
\ket{E} = \ket{E_0} + \frac{i}{\omega\epsilon_0} \sum_m \ket{E_m} \frac{\epsilon_i-\epsilon_b}{(\epsilon_m-\epsilon_i)(\epsilon_m-\epsilon_b)} \braket{E_m}{J},
\label{eq:Jepsform}
\end{equation}
yielding an expression of the form \eqref{eq:simpleJ}. The resulting \eqref{eq:Jepsform} is largely similar to \eqref{eq:E0sol}, but the integral $\braket{E_m}{J}$ is now no longer restricted to the interior of the inclusion, and receives contributions from all locations where $\bv{J}(\bv{r})$ is non-zero. Nevertheless, \eqref{eq:Jepsform} remains a rigorous solution of the Lippmann-Schwinger equation and still benefits from the completeness of the eigenmodes within the interior. 

Finally, the desired normal mode expansion of Green's tensor \eqref{eq:greenfn}, applicable to resonators in open and lossy systems, is obtained by choosing $\bv{J}(\bv{r})$ to be a localized Dirac-delta source. By the sifting property of Dirac-delta functions, the weight factor $\braket{E_m}{J}$ is simply the amplitude of the adjoint mode at the source location, $\bv{E}_m^\dagger(\bv{r})$. The Green's tensor is then constructed from its three components, giving
\begin{equation}
\tensor{G}(\bv{r},\bv{r}') = \tensor{G}_0(|\bv{r}-\bv{r}'|) + \frac{1}{k^2} \sum_m \frac{\epsilon_i-\epsilon_b}{(\epsilon_m-\epsilon_i)(\epsilon_m-\epsilon_b)} \bv{E}_m(\bv{r}) \otimes \bv{E}_m^\dagger(\bv{r}'),
\label{eq:greenexp}
\end{equation}
where $\tensor{G}_0(|\bv{r}-\bv{r}'|)$ is Green's tensor of the uniform background, \eqref{eq:greensfree}. The adjoint field $\bv{E}_m^\dagger(\bv{r}')$ is discussed in Appendix \ref{sec:adjoint}. Compared with \eqref{eq:simple}, three differences are immediately apparent: the switch from eigenfrequencies to eigenpermittivities, the extra term $\tensor{G}_0(|\bv{r}-\bv{r}'|)$, and the extra factor $(\epsilon_i-\epsilon_b)/(\epsilon_m-\epsilon_b)$. The extra term accounts for the divergence and longitudinal component of $\tensor{G}(\bv{r},\bv{r}')$ at $\bv{r}'$, while both the extra term and extra factor ensure the validity of \eqref{eq:greenexp} for all external sources, even though the eigenmodes only form a complete basis in the interior. Note that the extra factor vanishes at resonance, whereby \eqref{eq:greenexp} features the more usual weight $(\epsilon_m-\epsilon_i)^{-1}$. We demonstrate the importance of the extra term and factor in Sec.\ \ref{sec:numerics}. Finally, note that $\tensor{G}(\bv{r}, \bv{r}')$ becomes singular as $\bv{r}$ approaches $\bv{r}'$, and is longitudinal at $\bv{r}=\bv{r}'$. The term $\tensor{G}_0$ efficiently captures both aspects, even when expanding in terms of a small set of exclusively transverse modes. 

\section{Comparison with Quasinormal Modes}
\label{sec:quasi}
Quasinormal mode expansion, or resonant state expansion, has recently become a popular and successful generalization of \eqref{eq:simple} to lossy resonators in open systems. First introduced for quantum mechanical scattering in the context of nuclear physics,\autocite{siegert1939derivation}, they were later also introduced for electrodynamics problems.\autocite{baum1971singularity,baum1976emerging} Quasinormal modes are defined by the eigenvalue equation
\begin{equation}
\nabla\times(\nabla\times\bv{E}_m) - \frac{\omega_m^2}{c^2}\epsilon(\bv{r},\omega_m)\bv{E}_m = 0,
\label{eq:qeigen}
\end{equation}
which yields the Green's tensor expansion,\autocite{lee1999dyadic,doost2013resonant,kristensen2014modes,sauvan2014modal,muljarov2016exact}
\begin{equation}
\tensor{G}(\bv{r}, \bv{r}') = c^2\sum_m \frac{\bv{E}_m(\bv{r}) \otimes \bv{E}_m(\bv{r}')}{2\omega_m(\omega_m - \omega)}.
\label{eq:qgreen}
\end{equation}
A variation of \eqref{eq:qgreen} exists which is equivalent,\autocite{muljarov2016exact} as well as an alternative based on a scattered field representation.\autocite{bai2013efficient,ge2014quasinormal} The fundamental difference between quasinormal mode expansion and GENOME is the use of complex eigenfrequencies $\omega_m$ rather than complex eigenpermittivities $\epsilon_m$ to account for loss in lossy and open systems. However, this one change entails many ramifications, ranging from the fundamental to the practical. Due to the topical nature of quasinormal mode expansions, we devote this section to a thorough comparison of the advantages and disadvantages of the two methods.

\subsection{Completeness}
The completeness of quasinormal modes has been extensively discussed in the literature,\autocite{leung1994completeness,ching1998quasinormal,lee1999dyadic} being necessary for the validity of \eqref{eq:qgreen}. Rigorous proof shows that quasinormal modes form a complete set inside the inclusion, since a sharp interface at the inclusion boundary ensures that fields of sufficiently high spatial frequency are generated within via refraction to represent any field. However, the lack of sharp boundaries enclosing the background means that fields of sufficiently high spatial frequency are not generated.\autocite{leung1994completeness,ching1998quasinormal} Thus, \eqref{eq:qgreen} is rigorously valid only when $\bv{r}'$ is interior to the inclusion, but not when it is in the background. In contrast, GENOME \eqref{eq:greenexp} is complete for sources both inside and outside the inclusion, despite also employing modes \eqref{eq:eigen} that are complete only inside the inclusion. Specifically, GENOME contains an extra term and factor \eqref{eq:greenexp} relative to \eqref{eq:qgreen}, that can both be neglected on resonance. Hence, our expansion always converges to arbitrary accuracy regardless of detuning from resonance and distance from the resonator. We demonstrate this property with numerical examples in Sec.\ \ref{sec:numerics}.

For many practical applications, the formal lack of completeness of quasinormal modes is not so consequential, especially for Purcell factor calculations. Thus, quasinormal expansion \eqref{eq:qgreen} has been demonstrated to provide excellent numerical agreement even for exterior sources in numerous cases when the response is dominated by a few resonances.\autocite{sauvan2013theory,ge2014quasinormal,ge2014design,sauvan2014modal,yang2015simple} It is also claimed that formal issues of completeness may be bypassed in some practical implementations, where infinite space is mapped onto a finite simulation domain, thus finding all necessary modes.\autocite{lalanne2017light} This logic can be contrasted with the arguments of Leung et al.\autocite{leung1994completeness,ching1998quasinormal} Nevertheless, unphysical abnormalities can emerge away from resonance, such as negative Purcell factors.\autocite{sauvan2013theory,bai2013efficient} Meanwhile, other applications have more stringent requirements for completeness. For example, known modes of simple structures can be used to generate modes of a more complex structures, a process described in more detail in Section \ref{sec:numerics}.

A key advantage of modal expansions that employ either quasinormal or eigenpermittivity modes is that they treat open systems using only a discrete set of modes, avoiding the cumbersome continuum of radiation modes. Unfortunately, this advantage of quasinormal modes is lost when treating certain geometries, such as 2D structures or 3D resonators mounted on substrates.\autocite{doost2013resonant,vial2014quasimodal,lalanne2017light} In addition to the usual discrete set of modes, a continuous set of modes now emerges from the eigenvalue equation \eqref{eq:qeigen}, to account for a branch cut in the underlying dispersion relation. These transform the expansion \eqref{eq:qgreen} into an expression involving both a sum and an integral. Neglecting the continuum can lead to significant errors, especially for subwavelength resonators. However, when included, the continuum spoils the simplicity of quasinormal mode expansion and the analytic insights it offers. In contrast, we have demonstrated in the context of 2D dispersion relations that our eigenpermittivity modes do not experience any branch cuts, and a discrete set emerges from our eigenvalue equation \eqref{eq:eigen}.\autocite{chen2017robust}

Lastly, our generalized normal modes are always biorthogonal, and projection and expansion always proceeds via \eqref{eq:greenexp}, so the relative contributions modal contributions remain obvious from the detuning. Meanwhile, the biorthogonality normally enjoyed by quasinormal modes is disrupted by material dispersion, which can complicate some projection procedures.\autocite{sauvan2013theory,sauvan2014modal} The simple \eqref{eq:qgreen} no longer applies when more than one mode provides a significant contribution, and a more elaborate projection procedure is necessary. This involves evaluating overlap integrals between the modes and then inverting a linear system of equations.\autocite{sauvan2013theory,sauvan2014modal} This issue can be avoided by expansion via the Mittag-Leffler theorem, which yields \eqref{eq:qgreen} for interior sources even in the presence of dispersive $\epsilon(\bv{r},\omega)$.\autocite{leung1994completenessa}

\subsection{Far fields and normalization}
A hallmark peculiarity of quasinormal modes is their far-field behavior, diverging exponentially as $\bv{r}\rightarrow\infty$. This unphysical behavior is an unavoidable consequence of complex eigenfrequencies, in conjunction with radiating boundary conditions.\autocite{lai1990time} Divergence is imperceptible near the resonator, but dominates at further distances, with the divergence becoming noticeable earlier as $\imag(\omega_m)$ increases. Both source and detector coordinates of $\tensor{G}(\bv{r},\bv{r}')$ are affected, producing incorrect Purcell factors and radiation patterns, for example.\autocite{ge2014quasinormal} A remedy for quasinormal modes exists by numerically generating regularized modes from the original modes, via the Lippmann-Schwinger equation.\autocite{ge2014quasinormal} However, this assumes the response is dominated by a single mode, neglecting all other modes. Meanwhile, our eigenpermittivity modes are valid over the entire domain, both satisfying radiating boundary conditions and decaying to zero as $\bv{r}\rightarrow\infty$. This is the advantage of using eigenmodes of the Lippmann-Schwinger equation itself, defined at real frequencies.

The divergence of quasinormal modes complicated normalization, since any integral over the entire domain correspondingly diverges. Normalization is necessary for projection, so quasinormal modes were previously unsuitable for quantitative studies. More recently, pioneering efforts by several groups have led to several successful normalization schemes. Firstly, the diverging volume integral during normalization can be counterbalanced by a surface integral.\autocite{lai1990time,kristensen2012generalized,doost2013resonant,muljarov2016exact} However, some care is required when positioning the surface integral to avoid numerical sensitivity issues, particularly when the approximate form is used.\autocite{muljarov2016exact,kristensen2015normalization} The procedure is also inapplicable for backgrounds with non-uniform permittivity $\epsilon(\bv{r},\omega)$. Next, divergence can be quelled by perfectly matched layers along the simulation domain boundary, yielding a finite normalization integral.\autocite{sauvan2013theory} But this method can be inconvenient or even impossible to use unless a perfectly matched layer is used during simulation.\autocite{kristensen2015normalization} For example, resonators coupled to waveguides or periodic boundary conditions require separate treatment.\autocite{kristensen2014calculation} Estimates of the normalization constant can also be reverse-engineered by comparing the quasinormal mode expansion with a separate simulation using a test source, though this approximation is valid only for individual isolated resonances.\autocite{bai2013efficient} 

In comparison, a simple, general, and robust method exists to normalize our generalized normal modes, achieved by simply integrating the modes over the interior of the inclusion. Indeed, in our COMSOL implementation, we evaluate the necessary overlap integral of the mode with itself using simply the built-in volume integration tool.

\subsection{Other comparisons}
A vital practical consideration is the ease of generating the eigenmodes. Generalized normal modes are always defined by a single linear eigenvalue problem \eqref{eq:eigen}, and rapid robust solution is possible via the many powerful linear algebra packages. Furthermore, immediate implementation across a range of off-the-shelf simulation packages is possible. Quasinormal modes are relatively more difficult to find numerically since they are defined by a non-linear eigenvalue problem \eqref{eq:qeigen}. The non-linearity originates from the boundary conditions of the open system, since the eigenfrequency explicitly enters the Silver-M\"{u}ller radiation condition.\autocite{kristensen2014modes} This can be avoided in many numerical implementation through use of perfectly matched layers. An additional non-linearity is introduced when material dispersion is present, since frequency appears within $\epsilon(\bv{r}, \omega)$ of \eqref{eq:qeigen}. This non-linearity can also be eliminated by introducing auxilliary fields,\autocite{raman2010photonic} but this requires additional implementational effort and increases the number of equations to be solved. 

To handle the non-linearity, solutions for quasinormal modes commonly rely on a complex root search, requiring numerical iteration of \eqref{eq:qgreen} until a resonance is located. Accurate initial guesses of the eigenfrequencies are paramount, and it is impossible to guarantee that all relevant solutions have been found. These difficulties can be ameliorated by first analyzing the scattering spectrum of the target nanostructure, obtained through a separate numerical simulation using a broadband source. The real and imaginary parts of $\omega_m$ can be estimated from the peaks and linewidths of the spectrum. However, dark modes, low quality factor modes, and members of closely spaced resonances can remain elusive.\autocite{}

Finally, quasinormal modes create interpretational issues because the inclusion permittivity $\epsilon(\bv{r}, \omega)$ must be defined for complex frequencies, which is more awkward to interpret than real frequencies. Furthermore, the search for quasinormal modes cannot proceed unless $\epsilon(\bv{r}, \omega)$ is a sufficiently smooth function of frequency. This can preclude use of tabulated or experimentally measured permittivity data. Hence, approximations such as the Drude model are necessary, which must be extended to complex frequencies by analytic continuation. 

The preceeding discussion and subsections survey the advantages of GENOME over quasinormal mode expansion. Currently, quasinormal modes hold one key advantage over generalized normal modes: \eqref{eq:qgreen} is valid for all frequencies,\autocite{muljarov2016exact} so long as all aforementioned provisos are heeded. An approximate analytical expression can also be derived for the lineshapes of lossy resonators when a single quasinormal mode is dominant.\autocite{sauvan2013theory,sauvan2014modal} This enables great utility, as lineshapes are often the most accessible quantity in experiments. In contrast, generalized normal modes are defined for an individual frequency. To obtain expansions \eqref{eq:greenexp} at other frequencies, a new set of eigenmodes must be generated for each frequency. This represents the sacrifice necessary at present for a rigorous, robust eigenmode expansion of Green's tensor in an open, lossy system. 

However, this limitation of generalized normal modes is not fundamental. Firstly, once the eigenmodes have been obtained at one frequency, its eigenpermittivities at neighboring frequencies can be obtained perturbatively. To first order, the eigenmodes are also valid for a range of frequencies.\autocite{agranovich1999generalized} This allows the detuning factor in \eqref{eq:greenexp} to be reexpressed in terms of frequencies.\autocite{agranovich1999generalized} Furthermore, since the eigenmodes are complete, modes defined for one frequency are always able to represent modes at any other frequency via linear combination. Correspondingly, the variation of eigenpermittivities with frequency $\epsilon_m(\omega)$ can be obtained to arbitrary precision. Work is currently underway.

\section{Implementation and Numerical Examples}
\label{sec:numerics}
In practical terms, the key step in using GENOME is finding the eigenmodes and their eigenvalues. In this section, we describe some possible numerical implementations for the eigenmodes, before proceeding to detail our COMSOL Multiphysics implementation. We present some example eigenpermittivity modes produced by COMSOL. We then use these eigenmodes to expand the Green's tensor of the structure, thereby demonstrating GENOME. Finally, we compare GENOME with an incomplete naive modal expansion that also uses eigenpermittivity modes, demonstrating that only GENOME reproduces the real and imaginary parts of the Green's tensor correctly.

The linear eigenvalue equation for our modes is defined by \eqref{eq:eigen} in integral form. Since the kernel of \eqref{eq:eigen} is the Green's tensor of a uniform medium, the integral takes the form of a convolution, so efficient Fourier domain solutions are possible. However, the eigenmode equation need not be solved in integral form, and the differential form can be used instead,
\begin{equation}
\nabla\times(\nabla\times\bv{E}_m) - \epsilon_b k^2 \bv{E}_m = \frac{1}{s_m} \theta(\bv{r}) k^2 \bv{E}_m,
\label{eq:eigendiff}
\end{equation}
obtained from \eqref{eq:inhomowave} by setting $\bv{J} = 0$. This is the form we exploit for implementation on COMSOL. Additionally, simple structures such as spheres, slabs, and infinite cylinders admit analytic solutions via their well-known step-index dispersion relations.\autocite{bergman1980theory,farhi2016electromagnetic,chen2017robust} In particular, we have recently employed the argument principle method to the step-index fiber dispersion relation, enabling its efficient and robust solution. The supplied code can also easily be adapted to solve other transcendental equations.

Finally, \eqref{eq:eigen} can be efficiently solved for clusters of inclusions using the eigenmodes of its constituents as a basis. This exploits the fact that $\tensor{G}_0 (|\bv{r} - \bv{r}'|)$ is common to all inclusions, even if they have different shapes or compositions, while $\theta(\bv{r}')$ is non-zero only inside the inclusions.\autocite{bergman1980theory} The procedure then amounts to evaluating overlap integrals between known modes and diagonalizing a small dense matrix. Previous demonstrations include using modes of a sphere to generate modes of arbitrary clusters,\autocite{bergman1980theory} and periodic arrays.\autocite{bergman1979dielectrica,bergman1992bulk} Completeness is necessary to the success of this procedure. Note that this procedure bears similarities to ``perturbative'' methods, used in the context of quasinormal modes, which uses known modes of a simple inclusion to generate modes of any enclosed inclusion.\autocite{doost2014resonant} Completeness of quasinormal modes inside the inclusion ensures that the series always converges for ``perturbations'' of any depth. In this regard, our procedure is analagous, but is applicable to ``perturbations'' both within the inclusion and of the background.

\subsection{COMSOL Implementation}
COMSOL is a commercial numerical simulation package for the finite element method. We choose COMSOL for our eigenmode solver to demonstrate its ease of implementation on a widely used platform. COMSOL features an inbuilt eigenmode solver designed for complex eigenfrequency modes \eqref{eq:qeigen}. However, this solver is easily repurposed to solve for eigenpermittivity modes, re-expressed in differential form as \eqref{eq:eigendiff}. This is accomplished via a substitution trick, eventually allowing the eigenfrequencies found by COMSOL to be reinterpreted as $s_m$. Furthermore, the eigenmodes found by COMSOL are the true eigenpermittivity modes.

An adaptive mesh with finer resolution near the surface is desirable. This enables greater accuracy when finding the plasmonic modes, which have evanescent fields. It is also beneficial to enclose the simulation domain with perfectly matched layers, but this is not necessary since our generalized normal modes are well behaved at infinity. We thus forgo perfectly matched layers in generating the numerical examples below, despite being easy to implement in COMSOL, choosing instead to employ the inbuilt scattering boundary conditions to minimize backscattering from the simulation boundary.

Once found, the eigenmodes $\bv{E}_m$ require normalization according to \eqref{eq:norm} before use. This is accomplished by numerically evaluating $\bv{E}_m\cdot\bv{E}_m$ across the interior of the inclusion using the inbuilt integration function, yielding the normalization constant. As discussed in the Appendix, this integral can evaluate to zero for symmetry reasons, especially for analytic eigenmodes constructed using cylindrical or spherical harmonics. While this can be easily remedied, it was unnecessary since in our experience COMSOL already generates modes with non-zero normalization integrals, even for highly symmetric inclusions, such as cylinders. This is true even if no attempt is made to incorporate the symmetry of the inclusion into the simulation using symmetric boundary conditions.

With the eigenmodes in hand, use of GENOME \eqref{eq:greenexp} proceeds by evaluating the detuning factors and adding the known $\tensor{G}_0(|\bv{r}-\bv{r}'|)$ term for a point source, finally plotting the fields as desired. These steps were performed using MATLAB, interfacing with COMSOL via LiveLink.

\subsection{Numerical Examples}
\begin{figure}[!t]
\begin{center}
\includegraphics{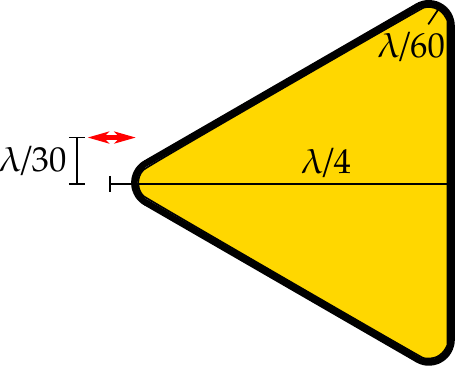}
\caption{A schematic of the simulated inclusion geometry, an equilateral triangle with rounded corners. Its height is $\lambda/4$ from the base to the imaginary unrounded apex, while the radius of the rounded segments is $\lambda/60$. The dipole source is offset by $\lambda/30$ from the imaginary apex, with the indicated orientation.}
\label{fig:trifig}
\end{center}
\end{figure}

\begin{figure}[!t]
\begin{center}
\subfloat[$\epsilon_m = -2.7-1.3i$]{\begin{tabular}[b]{c}%
	\includegraphics{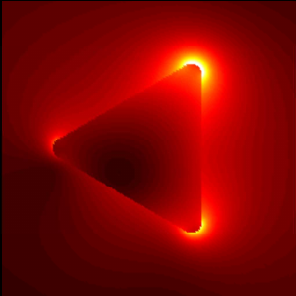}\\
	\includegraphics{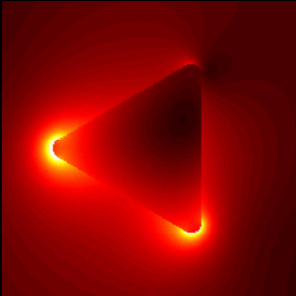}
	\end{tabular}}
\subfloat[$\epsilon_m = -0.46-0.13i$]{\begin{tabular}[b]{c}%
	\includegraphics{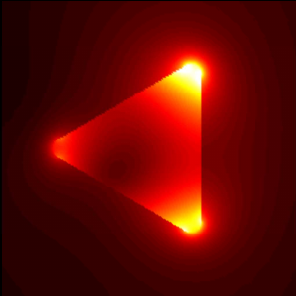}\\
	\includegraphics{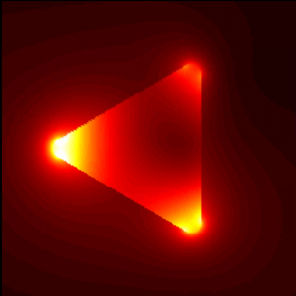}
	\end{tabular}}
\subfloat[$\epsilon_m = -0.87-0.04i$]{\begin{tabular}[b]{c}%
	\includegraphics{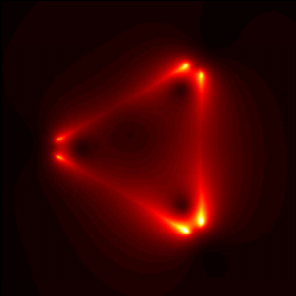}\\
	\includegraphics{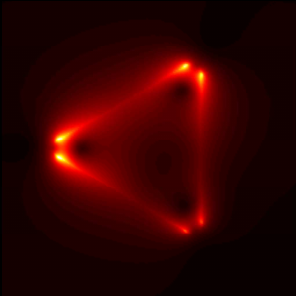}
	\end{tabular}}
\subfloat[$\epsilon_m = 2.4-2.7i,\newline\epsilon_m = 11-2.5i$]{\begin{tabular}[b]{c}%
	\includegraphics{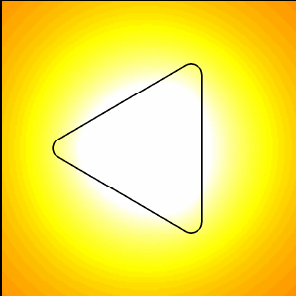}\\
	\includegraphics{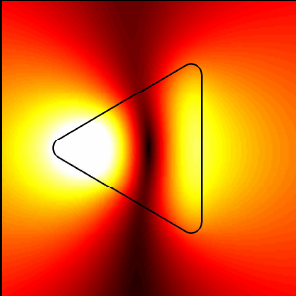}
	\end{tabular}}
\caption{(a)-(c) Each subfigure shows a degenerate member of a pair of plasmonic modes, along with their common eigenpermittivity. Since the fields of these modes are concentrated along the inclusion interface, the location of this interface well delineated by the fields themselves. (d) Shows the first two dielectric modes, with their respective eigenpermittivities. Here, we superimpose an outline of the triangular inclusion. All plots show $|E|$.}
\label{fig:modes}
\end{center}
\end{figure}
Firstly, we generated a set of modes using COMSOL for a two dimensional circular cylinder and checked them against analytic modes obtained from the step-index fiber dispersion relation.\autocite{chen2017robust} Both the eigenpermittivities and eigenmodal fields agree. Furthermore, we confirmed GENOME, \eqref{eq:greenexp}, using these COMSOL modes, comparing the results against direct calculation using cylindrical Mie theory, expanding the fields of a point source using Graf's addition theorem. Agreement was obtained, improving as more angular modes were included. This was limited by the numerical accuracy of the COMSOL modes generated, which suggests that there is no inherent accuracy limit to GENOME itself, for example due to incompleteness.

We proceed to demonstrate our COMSOL implementation of GENOME using a simple but non-trivial geometry, a two dimensional triangular inclusion shown in Figure \ref{fig:trifig}. Its corners have been rounded to avoid unphysical fields that would otherwise arise from geometrical singularities. The inclusion size is $\lambda/4$, which is subwavelength, but not small enough such that an electrostatic treatment would suffice. This choice serves to showcase GENOME as an electrodynamic tool. Note that choosing the inclusion size is equivalent to fixing the frequency of operation, since material dispersion is irrelevant to eigenpermittivity modes, which separate geometric properties from material properties.

Two types of symmetry are relevant to our chosen structure. Firstly, the two dimensional nature of the geometry separates all modes into strictly transverse magnetic or transverse electric polarizations. Secondly, the inclusion belongs to the $C_{3v}$ point group, which has three irreducible representations, $A_1$, $A_2$, and $E$, of which $E$ has a two-fold degeneracy. In generating the eigenmodes, we choose not to incorporate any of these symmetries into the simulation. Nevertheless, COMSOL produces eigenmodes with either in-plane or out-of-plane fields that are zero to within numerical noise. Furthermore, COMSOL naturally recognizes modes of degenerate pairs when a fine enough mesh is used.

We now present some representative eigenpermittivity modes found by COMSOL. We categorize all modes into two types. The first we denote as plasmonic modes, typically with eigenpermittivities $\real(\epsilon_m) < 0$. These have evanescent fields concentrated along the inclusion surface. The second type are dielectric modes, with $\real(\epsilon_m) > 0$, and field distributions more typical of a finite potential well. There are an infinite number of both types of modes.

The lowest order plasmonic modes are shown in Figure \ref{fig:modes} (a)--(c), along with their eigenpermittivities. Note that $\imag(\epsilon_m)$ of the fundamental plasmonic mode is larger than the higher order modes, indicating that it is a bright mode and the others are dark. All plasmonic modes shown are members of the $E$ representation of the $C_{3v}$ group, and are thus excited by in-plane dipole moments. Higher order plasmonic modes feature progressively more nodes along the surface of the inclusion. Including more modes within GENOME gives better quantitative agreement, especially for expanding the real part of the Green's function, and particularly when the source point is close to the surface. Progressively higher order modes have eigenpermittivities that asymptotically approach $\epsilon_m = -1$, which represents an accumulation point for the eigenvalue equation \eqref{eq:eigen}.\autocite{farhi2016electromagnetic} In particular, the imaginary parts of their eigenpermittivities become infinitesimal, indicating that they are non-radiative. Their contributions to GENOME also become increasingly negligible because their fields are increasing confined to the inclusion surface, and so have little overlap with any excitation.

The first two dielectric modes are also shown in Figure \ref{fig:modes} (d). The first mode belongs to the $A_1$ representation, and is thus excited by an out-of-plane dipole moment. The second mode belongs to the $E$ representation, and is a member of a degenerate pair. Higher order dielectric modes have more nodes, both in the radial and azimuthal directions. They also have larger positive $\real(\epsilon_m)$, so their contributions to GENOME also become increasingly small as they become detuned from any realistic inclusion permittivity.

\begin{figure}[!t]
\begin{center}
\subfloat[$\real(G_{xx})$]{\begin{tabular}[b]{c}%
	\includegraphics{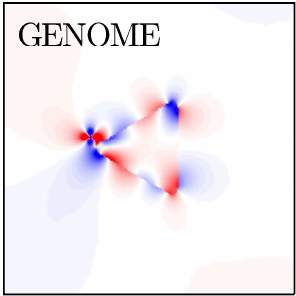}\\
	\includegraphics{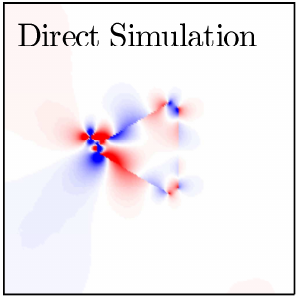}\\
	\includegraphics{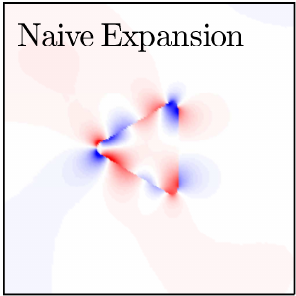}
	\end{tabular}}
\subfloat[$\real(G_{yx})$]{\begin{tabular}[b]{c}%
	\includegraphics{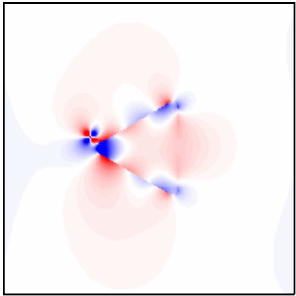}\\
	\includegraphics{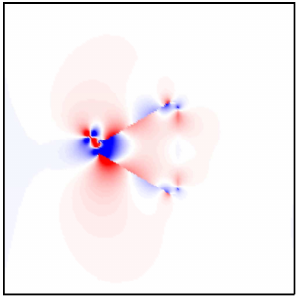}\\
	\includegraphics{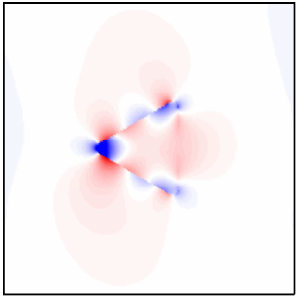}
	\end{tabular}}
\subfloat[$\imag(G_{xx})$]{\begin{tabular}[b]{c}%
	\includegraphics{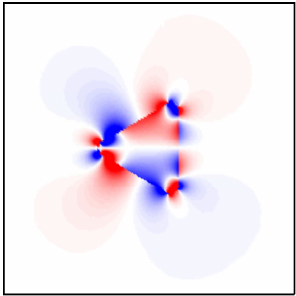}\\
	\includegraphics{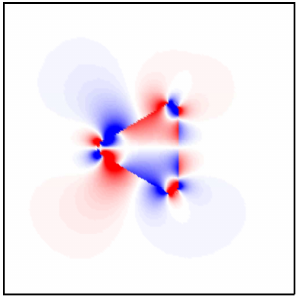}\\
	\includegraphics{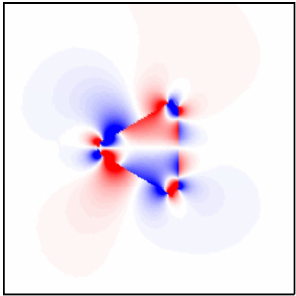}
	\end{tabular}}
\subfloat[$\imag(G_{yx})$]{\begin{tabular}[b]{c}%
	\includegraphics{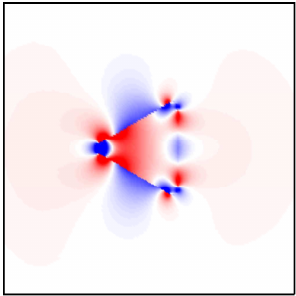}\\
	\includegraphics{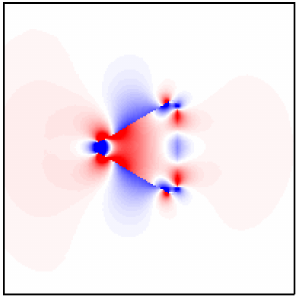}\\
	\includegraphics{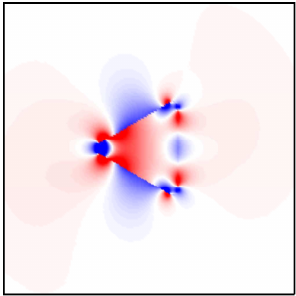}
	\end{tabular}}
\caption{Compares three different simulation methods of an identical geometry, with an $x$-oriented dipole placed near a triangular inclusion of $\epsilon_i = 0.87$. The first row shows the results of GENOME, the second row shows direct COMSOL simulation with a point source, and the third row shows the naive expansion \eqref{eq:gnaive} derived in Section \ref{sec:naive}. The scaling is identical between each row. Each column displays a different component of Green's tensor, as labeled. Again, we omit the outline of the inclusion from Figure \ref{fig:modes} (d), as the location of the boundary is clear from the fields themselves.}
\label{fig:darkexp}
\end{center}
\end{figure}

\begin{figure}[!t]
\begin{center}
\subfloat[$\real(G_{xx})$]{\begin{tabular}[b]{c}%
	\includegraphics{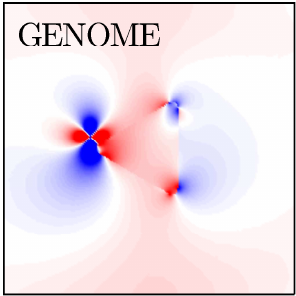}\\
	\includegraphics{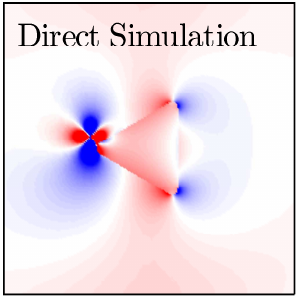}\\
	\includegraphics{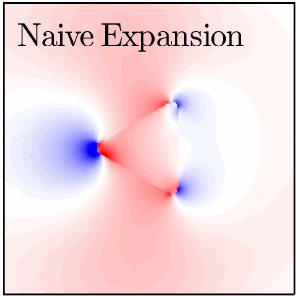}\\
	\end{tabular}}
\subfloat[$\real(G_{yx})$]{\begin{tabular}[b]{c}%
	\includegraphics{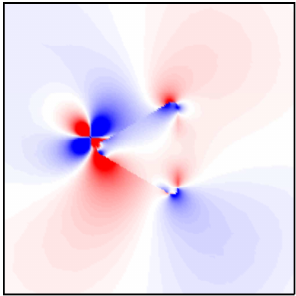}\\
	\includegraphics{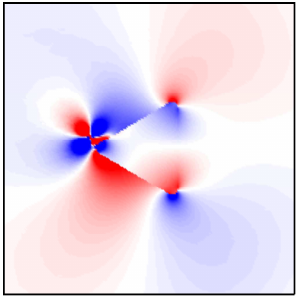}\\
	\includegraphics{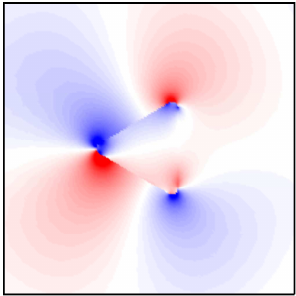}
	\end{tabular}}
\subfloat[$\imag(G_{xx})$]{\begin{tabular}[b]{c}%
	\includegraphics{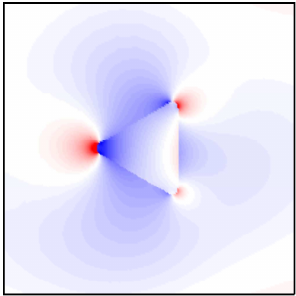}\\
	\includegraphics{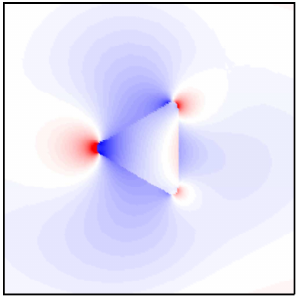}\\
	\includegraphics{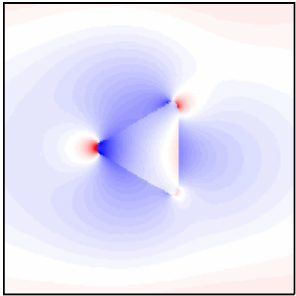}
	\end{tabular}}
\subfloat[$\imag(G_{yx})$]{\begin{tabular}[b]{c}%
	\includegraphics{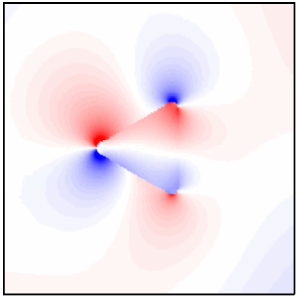}\\
	\includegraphics{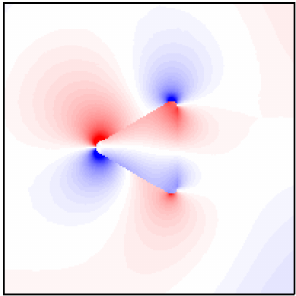}\\
	\includegraphics{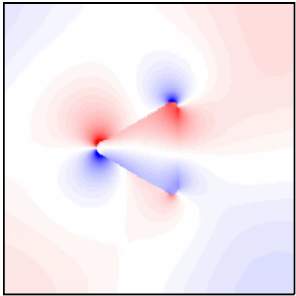}
	\end{tabular}}
\caption{As in Figure \ref{fig:darkexp}, but the permittivity of the triangular inclusion is $\epsilon_i = -2.7$.}
\label{fig:brightexp}
\end{center}
\end{figure}

We now demonstrate the use of these modes within GENOME via \eqref{eq:greenexp}. We provide three examples, with inclusion permittivities of $\epsilon_i=-0.87$, $\epsilon_i=-2.7$, and $\epsilon_i=-5.0$, all relying on the some modes. This places the inclusion ``on resonance'' with a dark plasmonic mode, with the bright plasmonic mode, or out of resonance with all modes, respectively. In order to produce plots, we choose to locate the source dipole near the leftmost tip of the triangular inclusion and orient its moment along $x$. The resulting $E_x$ and $E_y$ fields represent $G_{xx}$ and $G_{yx}$, and are plotted in Figures \ref{fig:darkexp}--\ref{fig:detunedexp}, showing both real and imaginary parts. To provide a benchmark for GENOME, we plot a direct simulation of a radiating dipole source produced by COMSOL without using any eigenmode expansion. Also plotted is the result of a naive expansion relevant to Section \ref{sec:naive}. Out-of-plane electric fields are not shown, as they are of the order $10^{-8}$, representing numerical noise.

Figure \ref{fig:darkexp} shows the simulation of the $\epsilon_i = -0.87$ inclusion. The dominant pair of modes excited is that of Figure \ref{fig:modes} (c), with eigenpermittivity $\epsilon_m = -0.87 - 0.04i$. The imaginary part, $\imag(\epsilon_m)$, is relatively small, so this resonance is relatively easy to approach with a passive medium,\footnote{In our notation, a passive medium has $\imag(\epsilon)>0$.} meaning a small detuing is often achievable. Thus, the contribution of this one pair of modes is dominant, being approximately 5 times greater than the next most dominant pair, seen in Figure \ref{fig:modes} (b). As demonstrated in Figure \ref{fig:darkexp}, GENOME obtains quantitative agreement with the benchmark direct simulation, reproducing both the real and imaginary parts of the Green's tensor. In particular, graphical accuracy is obtained for the imaginary part, while the divergence in the real part of the Green's tensor at the source location is also efficiently reproduced. Note that eight plasmonic modes were used in the expansion, representing four degenerate pairs.

The advantage of modal completeness is that good agreement is obtained whether the inclusion is on resonance or detuned from resonance, as demonstrated in our next example, in Figure \ref{fig:brightexp}. The inclusion permittivity is $\epsilon_i = -2.7$, which is closest to the pair of bright modes of Figure \ref{fig:modes} (a), with $\epsilon_m = -2.7 - 1.3i$. However, the radiative nature of this pair means that they have a relatively large negative $\imag(\epsilon_m)$, so its detuning from a passive medium can never be arbitrarily small. Consequently, the bright pair of modes is not dominant, and their contribution is roughly equal to the next dominant pair. Higher order dark modes also become more important. Here, the imaginary part of the Green's tensor is again reproduced to graphical accuracy, while the real part shows small quantitative disagreements due to the lack of higher order plasmonic modes. In our third example, Figure \ref{fig:detunedexp}, we show that GENOME continues to produce good agreement even for inclusions detuned from all resonances. Here, $\epsilon_i = -5.0$, but quantitative agreement is still achieved using the same four pairs of plasmonic modes.

\subsection{Comparison with a naive expansion}
\label{sec:naive}
\begin{figure}[!t]
\begin{center}
\subfloat[$\real(G_{xx})$]{\begin{tabular}[b]{c}%
	\includegraphics{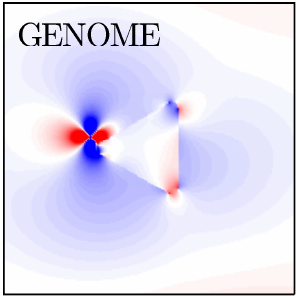}\\
	\includegraphics{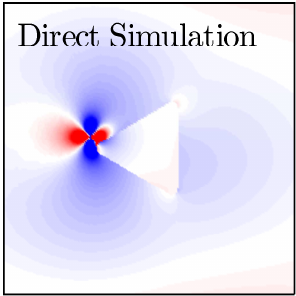}\\
	\includegraphics{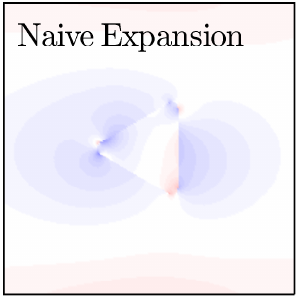}
	\end{tabular}}
\subfloat[$\real(G_{yx})$]{\begin{tabular}[b]{c}%
	\includegraphics{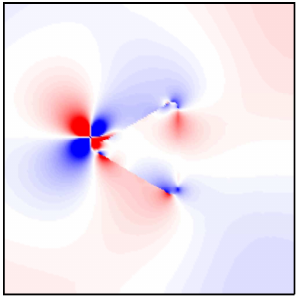}\\
	\includegraphics{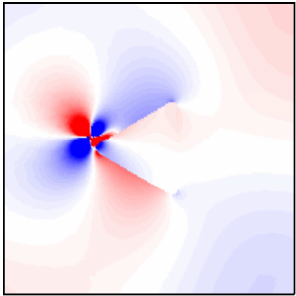}\\
	\includegraphics{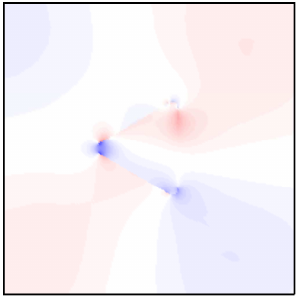}
	\end{tabular}}
\subfloat[$\imag(G_{xx})$]{\begin{tabular}[b]{c}%
	\includegraphics{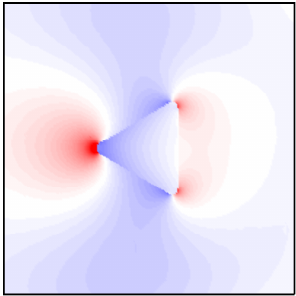}\\
	\includegraphics{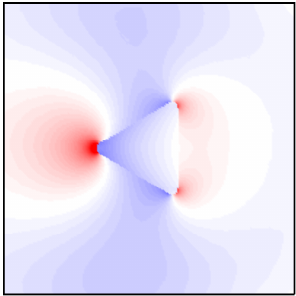}\\
	\includegraphics{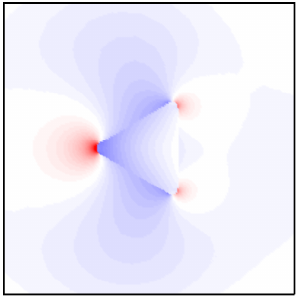}
	\end{tabular}}
\subfloat[$\imag(G_{yx})$]{\begin{tabular}[b]{c}%
	\includegraphics{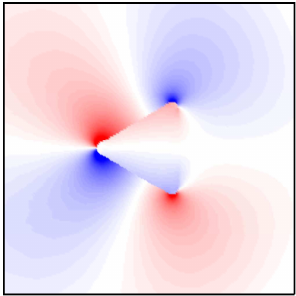}\\
	\includegraphics{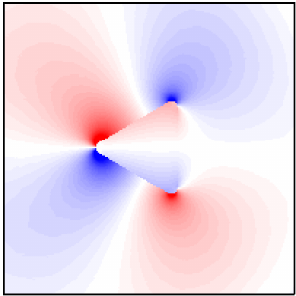}\\
	\includegraphics{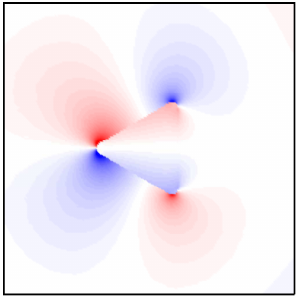}
	\end{tabular}}
\caption{As in Figure \ref{fig:darkexp}, but the permittivity of the triangular inclusion is $\epsilon_i = -5.0$.}
\label{fig:detunedexp}
\end{center}
\end{figure}
We now demonstrate the importance of completeness, and thus using the Lippmann-Schwinger equation in constructing GENOME. If instead we expand $\tensor{G}(\bv{r}, \bv{r}')$ directly in terms of our eigenpermittivity modes \eqref{eq:eigen}, with some coefficient $\bv{\alpha}_m(\bv{r}')$,
\begin{equation}
\tensor{G}(\bv{r}, \bv{r}') \approx \sum_m \bv{\alpha}_m(\bv{r'}) \otimes \bv{E}_m(\bv{r}),
\label{eq:naiveexp}
\end{equation}
we arrive at the false assumption that our eigenmodes are complete everywhere. Inserting \eqref{eq:naiveexp} into \eqref{eq:greenfn} and projecting using the biorthogonality of eigenpermittivity modes, we arrive at the naive, erroneous expansion
\begin{equation}
\tensor{G}(\bv{r},\bv{r}') \approx \frac{1}{k^2} \sum_m \frac{\bv{E}_m(\bv{r}) \otimes \bv{E}_m^\dagger(\bv{r}')}{\epsilon_m-\epsilon_i},
\label{eq:gnaive}
\end{equation}
which is not valid when the source $\bv{r}'$ is placed in the background. Analogous assumptions and procedures are sometimes employed by other expansion methods, e.g., when using quasinormal modes. 

In Figures \ref{fig:darkexp}--\ref{fig:detunedexp}, we compare the results of rigorous GENOME \eqref{eq:greenexp} with the naive expansion \eqref{eq:gnaive}, exposing the inaccuracy of \eqref{eq:gnaive}. Firstly, the naive expansion is always missing the divergence in the real part at the source. Satisfactorily agreement in the imaginary parts of the field is possible when the detuning is small, such as in Figure \ref{fig:darkexp}. But as detuning grows, even the agreement in the imaginary part begins to degrade. This can be seen in Figure \ref{fig:brightexp}, for an inclusion ``on resonance'' with the bright mode. Moving completely away from all resonances, any agreement continues to degrade, as seen in Figure \ref{fig:detunedexp}. In Figures \ref{fig:darkexp}--\ref{fig:detunedexp}, the same four pairs of modes used for GENOME were used in the expansion of \eqref{eq:gnaive}. It is important to note that the accuracy of \eqref{eq:gnaive} never improves when more modes are used, always converging to an incorrect result.

\section{Summary}
\label{sec:conc}
In this paper, we developed GENOME, a modal expansion for the electromagnetic Green's tensor based on stationary normal modes generalized to handle lossy resonators in open systems. Its foundation is the Lippmann-Schwinger equation, \eqref{eq:lippsch}, introduced in Section \ref{sec:lippsch}. Expansion proceeded using eigenmodes of the Lippmann-Schwinger equation, defined in Section \ref{sec:eigenmodes}. Crucially, we defined the permittivity of the inclusion as the eigenvalue, resulting in the linear eigenvalue equation \eqref{eq:eigen}. Physical interpretations of eigenpermittivity modes and some of their properties were also discussed in Section \ref{sec:eigenmodes}. The expansion itself was derived in Section \ref{sec:expand}, culminating in the final GENOME expression \eqref{eq:greenexp}.

Since GENOME employs true normal modes, it carries a number of advantages, as discussed in Section \ref{sec:quasi}. The modes remain discrete and are biorthogonal, which greatly facilitates modal expansion. The modes are complete, ensuring that the expansion always converges towards the true solution of the target inhomogeneous differential equation \eqref{eq:maxwell}. This remains true of the far-fields, where our modes intrinsically satisfy the governing source-free differential equation, thus simultaneously obeying Sommerfeld boundary conditions and decaying to zero. This has the additional benefit of trivializing normalization.

We described numerical implementation of GENOME in Section 4, focusing on the key step of generating the eigenpermittivity modes. Since the defining eigenvalue equation is linear, several possibilities were described. In particular, pre-existing eigenmode solvers can be adapted with a simple substitution trick, described in Section 4.1. In Section 4.2, we presented results from our COMSOL implementation, based on the differential form the defining eigenmode equation \eqref{eq:eigendiff}. We obtained the modes of a triangular inclusion, and provided some further discussion on the characteristics of eigenpermittivity modes. We then proceeded to use these modes in GENOME, comparing our expansion with a direct COMSOL simulation of a point source. Graphical accuracy in the radiated fields was obtained, particularly for the imaginary part. Finally, we demonstrated the importance of completeness, by comparing GENOME with a naive expansion that also uses eigenpermittivity modes, \eqref{eq:gnaive}, which never converges towards the true solution, regardless of the number of modes used.

\section*{Acknowledgments}
We would like to thank A. Farhi, P. Lallane, K. Vynck for many useful discussions. PYC and YS were partially supported by Israel Science Foundation (ISF) (899/16) and by the Israeli National Nanotechnology Initiative.

\appendix
\section{Adjoint modes}
\label{sec:adjoint}
We now give the explicit forms for the overlap integrals in \eqref{eq:E0sol},
\begin{equation}
\bra{E_m}\hat{\theta}\ket{E_0} = \int \theta(\bv{r}) \bv{E}_m^\dagger(\bv{r}) \cdot \bv{E}_0 (\bv{r})\, d\bv{r},
\label{eq:overlapE0}
\end{equation}
and in \eqref{eq:Jepsform}, 
\begin{equation}
\braket{E_m}{J} = \int \bv{E}_m^\dagger(\bv{r}) \cdot \bv{J}(\bv{r})\, d\bv{r}.
\label{eq:overlapJ}
\end{equation}
The adjoint field $\bv{E}_m^\dagger(\bv{r})$ in \eqref{eq:overlapE0} and \eqref{eq:overlapJ} is not necessarily the complex conjugate field $\bv{E}_m^*(\bv{r})$, which is the familiar form of $\bra{E_m}$ for a self-adjoint or Hermitean operator. Instead, the operator $\hat{\Gamma}\hat{\theta}$ in \eqref{eq:gammadef} is symmetric, so the adjoint field is identical to the direct field,\autocite{bergman1980theory}
\begin{equation}
\bv{E}_m^\dagger(\bv{r}) = \bv{E}_m(\bv{r}).
\label{eq:adjoint}
\end{equation}

\eqref{eq:adjoint} is true unless the structure itself possesses symmetry. For example, an infinite cylinder has both continuous translational symmetry and continuous rotational symmetry, giving rise to $e^{i\beta z}$ and $e^{im\theta}$ variations in the respective directions. In this case, \eqref{eq:adjoint} must be modified, and the adjoint field is obtained by the substitutions $\beta \rightarrow -\beta$ and $m \rightarrow -m$, while leaving the radial variation of the mode unchanged.\autocite{bergman1980theory,lai1990time,farhi2016electromagnetic,kristensen2015normalization} Alternatively, the modes may be constructed using sine and cosine linear combinations of $e^{\pm i\beta z}$ and $e^{\pm im\theta}$, and \eqref{eq:adjoint} once again becomes true.\autocite{lai1990time,doost2013resonant,kristensen2015normalization}

\printbibliography
\end{document}